\documentclass{cmspaper}

\usepackage{amssymb}
\usepackage[dvips]{graphicx}
\usepackage{rotating}
\usepackage{color}

\newcommand{\be}{\begin{eqnarray}}
\newcommand{\ee}{\end{eqnarray}}

\def\BE{\begin{equation}}
\def\EE{\end{equation}}

\def\lsim{\mathrel{\raise.3ex\hbox{$<$\kern-.75em\lower1ex\hbox{$\sim$}}}}
\def\gsim{\mathrel{\raise.3ex\hbox{$>$\kern-.75em\lower1ex\hbox{$\sim$}}}}

\def\fun#1#2{\lower3.6pt\vbox{\baselineskip0pt\lineskip.9pt}}
\makeatletter

\def\vereq#1#2{\lower3pt\vbox{\baselineskip1.5pt \lineskip1.5pt
\ialign{$\m@th#1\hfill##\hfil$\crcr#2\crcr\sim\crcr}}}

\makeatother
\newcommand{\jpsi}{$J/\psi$}
\newcommand{\psip}{$\psi^{\prime}$}
\newcommand{\ups}{$\Upsilon$}
\newcommand{\upsp}{$\Upsilon^{\prime}$}
\newcommand{\upspp}{$\Upsilon^{\prime\prime}$}
\newcommand{\ET}{E_T}

\newcommand{\smallurl}[1]{{\small{\url{#1}}}}

\newcommand{\dNdeta}{dN_{\rm ch}/d\eta|_{\eta=0}}

\begin{document}

%==============================================================================
% title page for few authors

\begin{titlepage}

% select one of the following and type in the proper number:
  \conferencereport{2006/066}
   \date{21 September 2006}

  \title{Hard Probe Capabilities of CMS in Heavy Ion Collisions at the LHC}

\begin{Authlist}
    Gunther Roland for the CMS Collaboration
       \Instfoot{MIT}{Massachusetts Institute of Technology, \\ 77 Massachusetts Ave, Cambridge, MA 02139 USA}
  \end{Authlist}

%\address{Massachusetts Institute of Technology, \\ 77 Massachusetts Ave, Cambridge, MA 02139 USA}

\begin{abstract}

Heavy ion collisions at the Large Hadron Collider (LHC) will produce strongly 
interacting matter at unprecedented energy densities. At LHC collision energies,
new hard probes of the dense initial collision system will become readily available.
We present an overview of the capabilities of the Compact Muon Solenoid (CMS) detector to use these probes 
for a detailed study of QCD phenomenology at the highest  energy densities.
\end{abstract}

\conference{Presented at {\it Hard Probes 2006}, Asilomar, June 15, 2006}
  
\end{titlepage}

\section{Overview}
Data collected by the four experiments at the Relativistic Heavy Ion Collider
(RHIC) suggest that in heavy ion collisions at
$\sqrt{s_{_{\it NN}}} =$ 200~GeV an equilibrated partonic
system is formed.  There is strong evidence that this dense medium is
highly interactive, perhaps best described as a quark-gluon liquid, and is
almost opaque to fast partons.  In addition, many surprisingly simple
empirical relationships describing the global characteristics of particle
production have been found \cite{WhitePapers}.  
The LHC will collide Pb ions at $\sqrt{s_{_{\it NN}}} =$ 5.5~TeV, the biggest step
in collision energy in the history of our field. 
Measurements at this energy
will either confirm and extend the theoretical picture emerging from RHIC
or challenge and redirect our understanding of strongly
interacting matter at extreme densities.  
Progress at the LHC will not only come from the increased initial energy density, but also
through a greatly expanded mass and $p_T$ range of hard probes, such as
high $p_T$ jets and photons, $Z^0$ bosons, the $\Upsilon$ states, $D$ and $B$ mesons, and
high-mass dileptons.
Below we describe the characteristics of the Compact Muon Solenoid (CMS) detector 
that will allow us to exploit the new opportunities presented by the LHC.

\section{CMS as a detector for heavy ion physics}\label{appar}

The discoveries at RHIC have not only transformed our picture of
nuclear matter at extreme densities, but have also  shifted the
emphasis in the observables best suited for extracting the
properties of the initial high-density QCD system~\cite{WhitePapers}. Examples of these
observables include elliptic flow, very high $p_T$
jets and heavy quarkonia. The importance of hard probes implies the  
need for detectors with large acceptance, high rate capability and high
resolution, leading to a
convergence of experimental techniques between heavy ion
and particle physics. Using CMS for heavy ion collisions takes this
development to its logical conclusion, leveraging the 
extensive resources that have already gone into the development
and construction of the apparatus. Below we discuss 
the extent to which CMS fulfills the criteria for an ideal heavy ion
detector.

\begin{figure}[hbtp]
\begin{center}
\includegraphics[width=10cm]{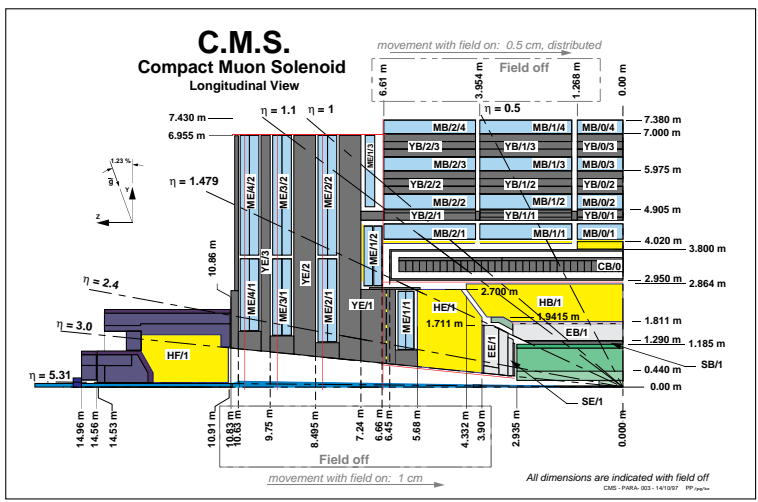}
\caption {A vertical slice through one quadrant of the CMS detector. The other
three quadrants are identical and symmetric about the center of the
interaction region and about the beam axis.}
\label{SideView}
\end{center}
\end{figure}

CMS was designed to provide tracking and calorimetry with high
resolution and granularity over the full azimuthal angle as well as over a very
large rapidity range. 
It is capable of precise detection of
muons, electrons, photons, jets, and heavy flavor tagging via 
secondary vertices.  The 
detector is symmetric about both the beam axis and the center of the nominal
interaction region.  The overall layout of the  apparatus is illustrated in
Fig.~\ref{SideView}.  A detailed description is provided in 
the Technical Design Reports~\cite{TDRs}.
Four of the requirements for a heavy ion detector at LHC for which CMS excels are discussed
below.

\begin{figure}[ht]
\centering
\includegraphics[width=10cm]{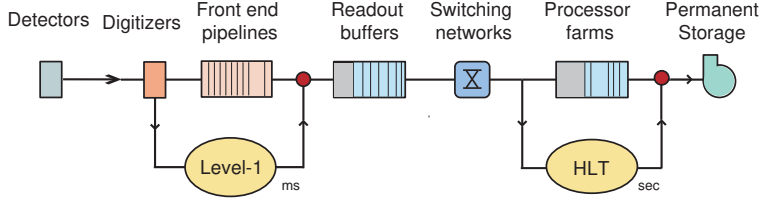}
\caption{Schematic of the data flow through the CMS DAQ system.}
\label{fig:data_flow}
\end{figure}

  \textbf{High rate}: The CMS DAQ and trigger system (see Fig.~\ref{fig:data_flow})
  is designed to deal with p+p collisions at
  luminosities of up to $10^{34}$~cm$^{-2}$s$^{-1}$, corresponding to p+p
  event rates of 40~MHz. As a result, the fast detector technologies chosen
  for tracking (Si-pixels and strips), electromagnetic and hadronic
  calorimetry, and  muon identification means that, for Pb+Pb collisions, CMS can be read out
  with a minimum bias trigger at the full expected
  luminosity. This fast readout allows detailed inspection of
  every event in the High Level Trigger (HLT) farm. The HLT CPU resources
are sufficient to run complex analysis algorithms on each 
  event, making a complete selection and archiving of events
  containing rare probes, such as extremely high $p_{T}$ jets or 
  high mass dileptons, possible.

\begin{figure}[htb]
 \includegraphics[width=0.32\textwidth]{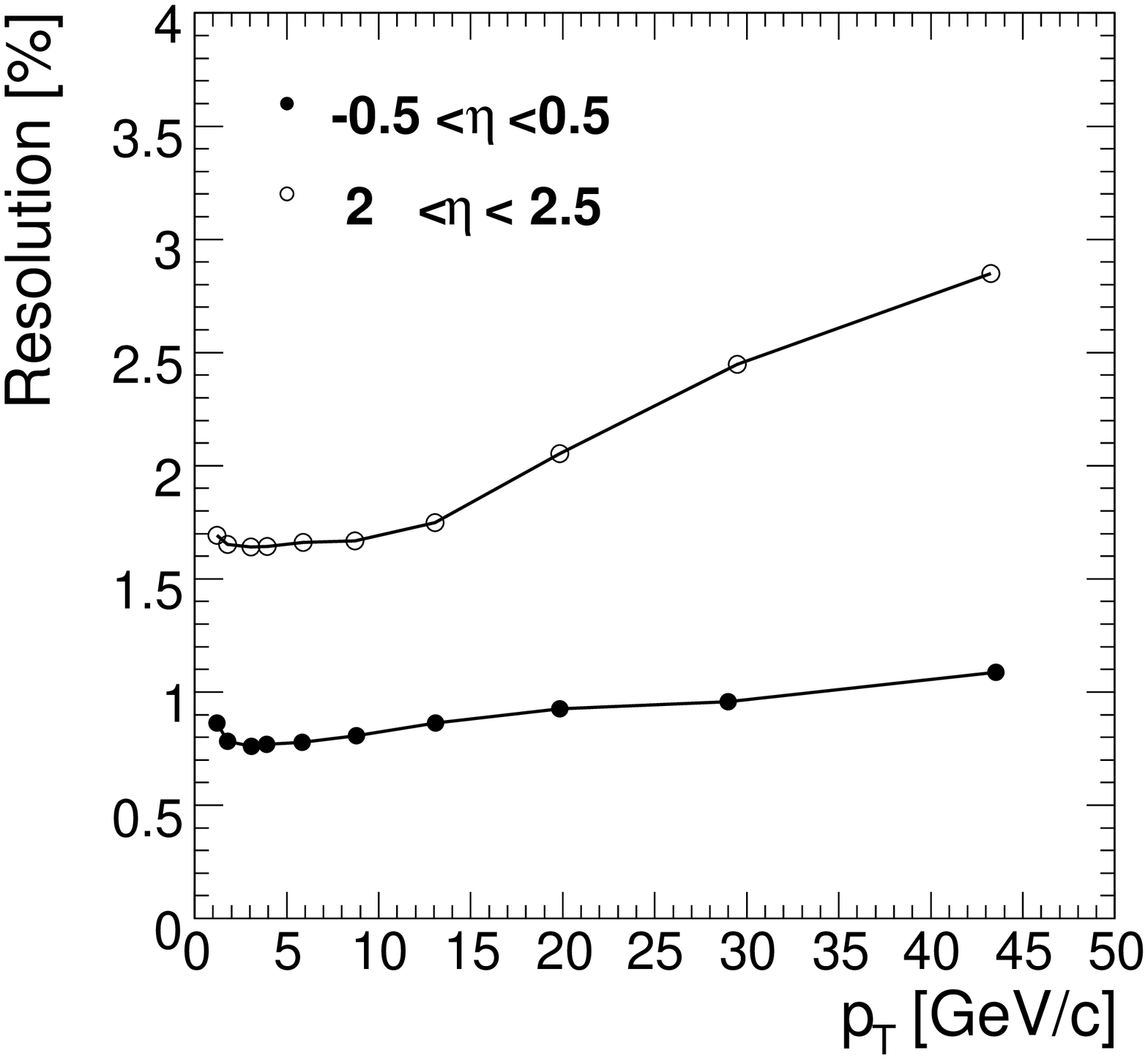}
 \includegraphics[width=0.32\textwidth]{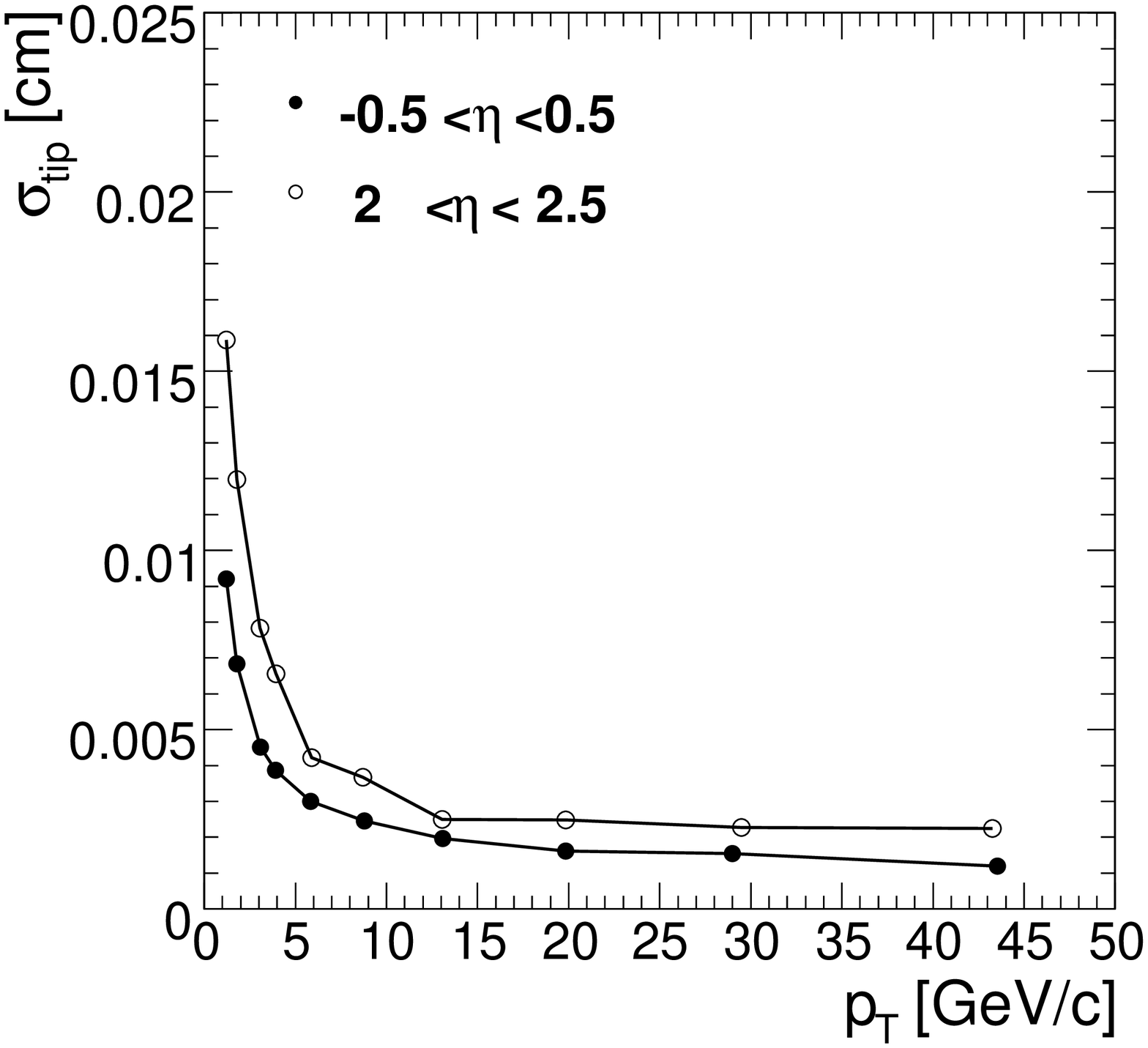}
 \includegraphics[width=0.32\textwidth]{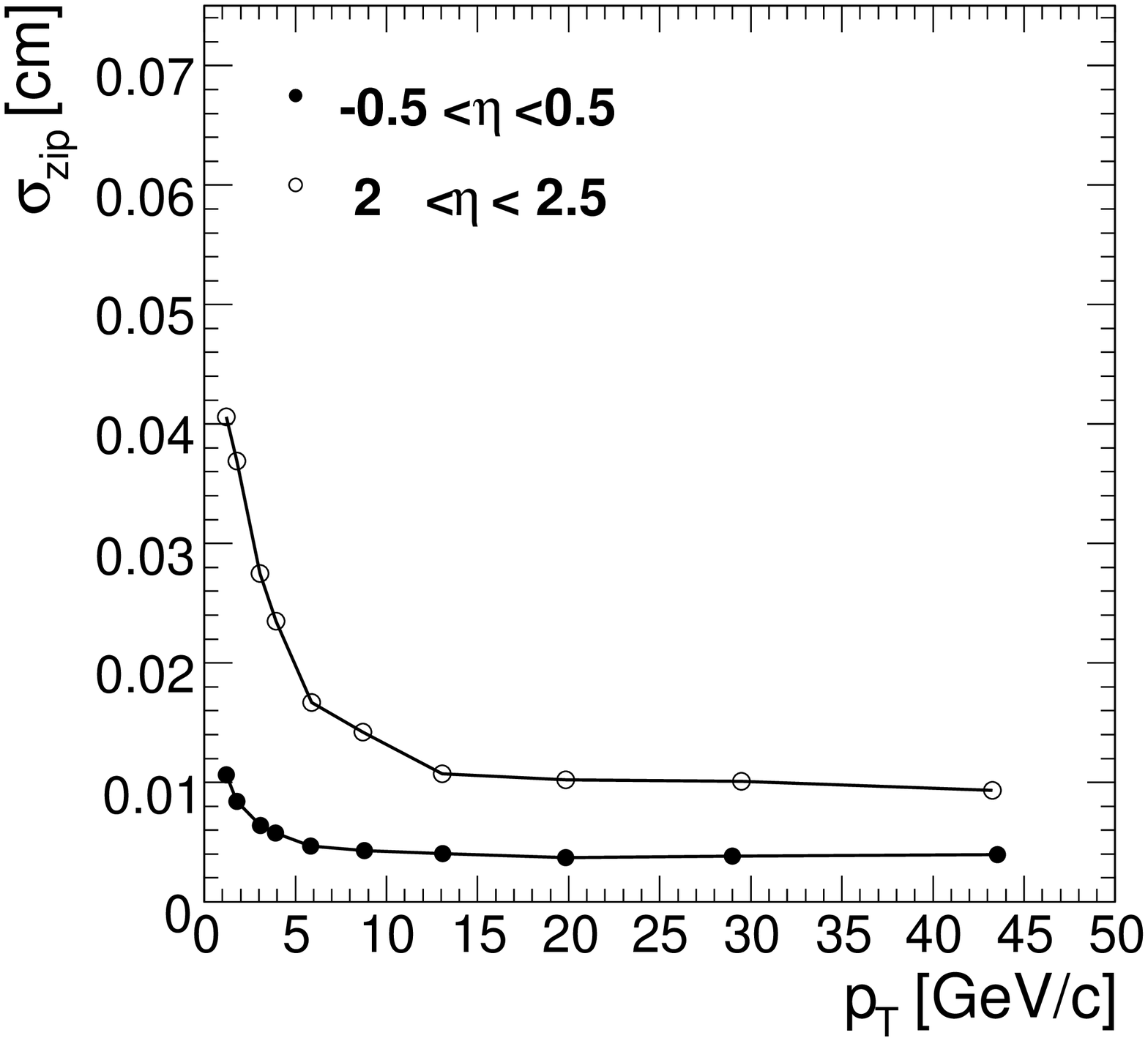}
\caption{The $p_T$ dependence of the track parameter resolution
achieved in heavy ion events in the barrel region (full symbols) and
in the forward region (open symbols). (Left) Transverse momentum
re\-so\-lu\-tion. (Center) Transverse track-pointing
re\-so\-lu\-tion. (Right) Longitudinal track-pointing
re\-so\-lu\-tion. } \label{fig:reso}
\end{figure}

  \textbf{High resolution and granularity}:
  At the full p+p luminosity there will be, on average, 20
  collisions per bunch crossing. To disentangle very
  high momentum observables with $p_{T}>500$~GeV/c in this
  environment, the resolution and granularity of all detector
  components has been pushed to the extreme, consequently making the detector
  well  suited to the high multiplicity conditions in central heavy ion collisions.
  The high granularity of the Si-pixel layers, in combination with
  the 4~T magnetic field, results in the world's best momentum resolutioni for charged particle tracks,
  $\Delta p_{T}/p_{T} < 1.5\%$ up to $p_{T}\approx 100$~GeV/c. At
  the same time, a track-pointing resolution of better than 50~$\mu$m
  (less than $20$~$\mu$m for $p_{T}>10$~GeV/c) is achieved. For
  $dN_{\rm ch}/dy\approx 3000$ in Pb+Pb collisions, tracks can be reconstructed
  with an efficiency of $\sim 80\%$ with a low rate of of fake tracks.
The momentum and track-pointing resolution
achieved in heavy ion collisions (see Fig.~\ref{fig:reso}) are
comparable to those in the low occupancy p+p events.

\begin{figure}[htb]
\centering
\includegraphics[width=10cm]{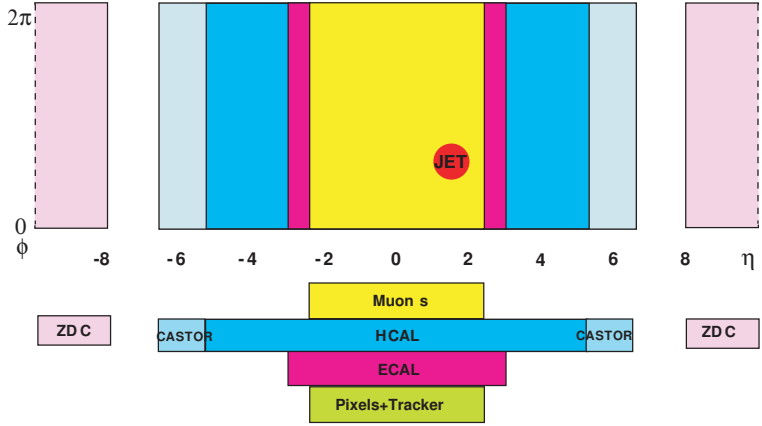}
\caption{Acceptance of tracking, calorimetry, and muon
identification in pseudorapidity and azimuth. The size of a jet with
cone $R=0.5$ is also depicted as an illustration.}
\label{fig:coverage}
\end{figure}

\textbf{Large acceptance tracking and calorimetry}: CMS includes high resolution tracking and calorimetry over 
  2$\pi$ in azimuth and a
  uniquely large range in rapidity. The
  acceptance of the tracking detectors, calorimeters and muon
  chambers can be seen in Fig.~\ref{fig:coverage}.  The Zero Degree
  Calorimeters ($|\eta_{\rm neutral}| >8.0$) and the CASTOR detector
  ($5.2<|\eta|<6.6$) will allow measurements of low-$x$ phenomena and
  particle and energy flow at very forward rapidities.

\begin{figure}[htb]
\centering
  \includegraphics[width=0.404\textwidth]{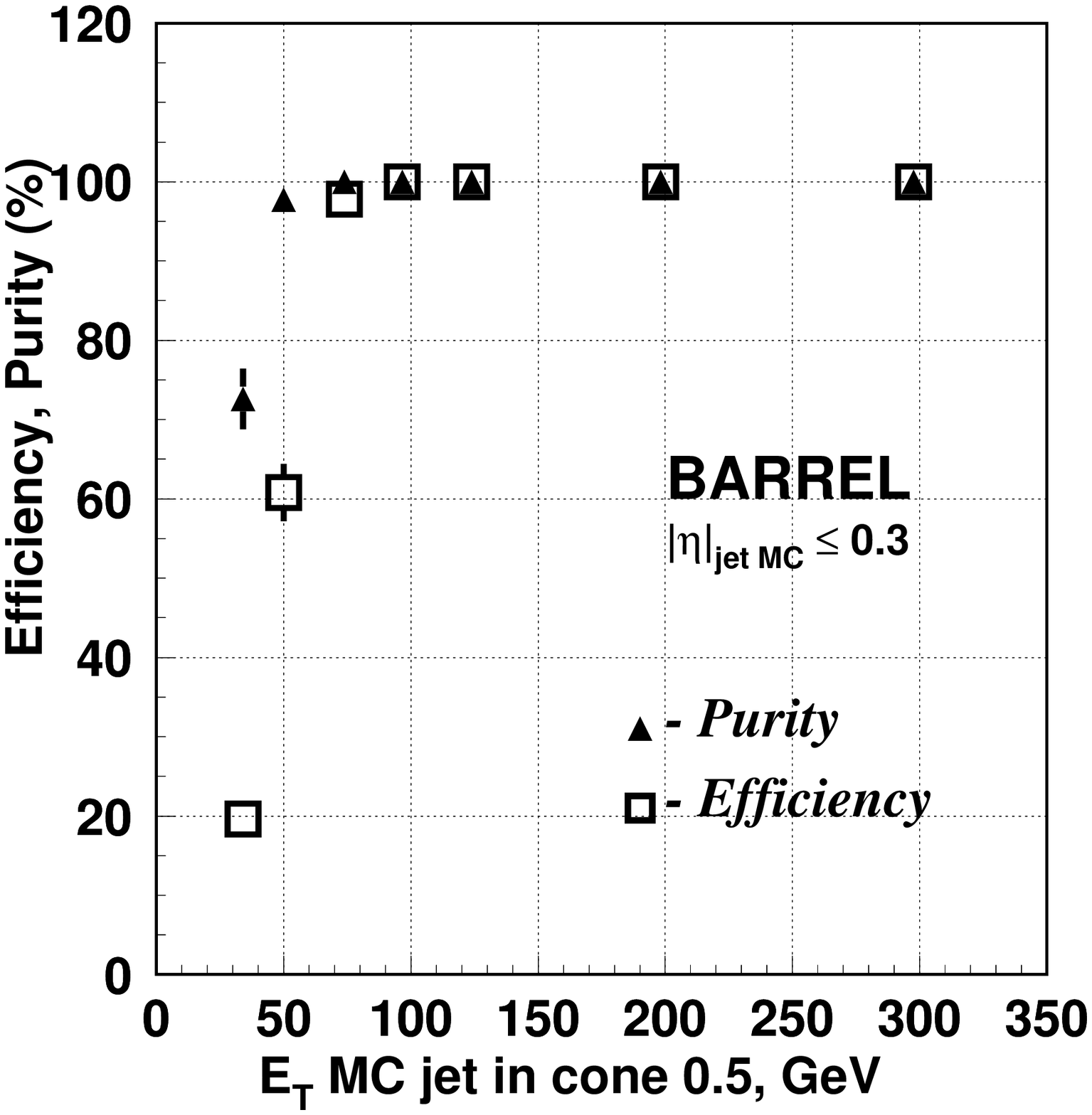}
  \includegraphics[width=0.537\textwidth]{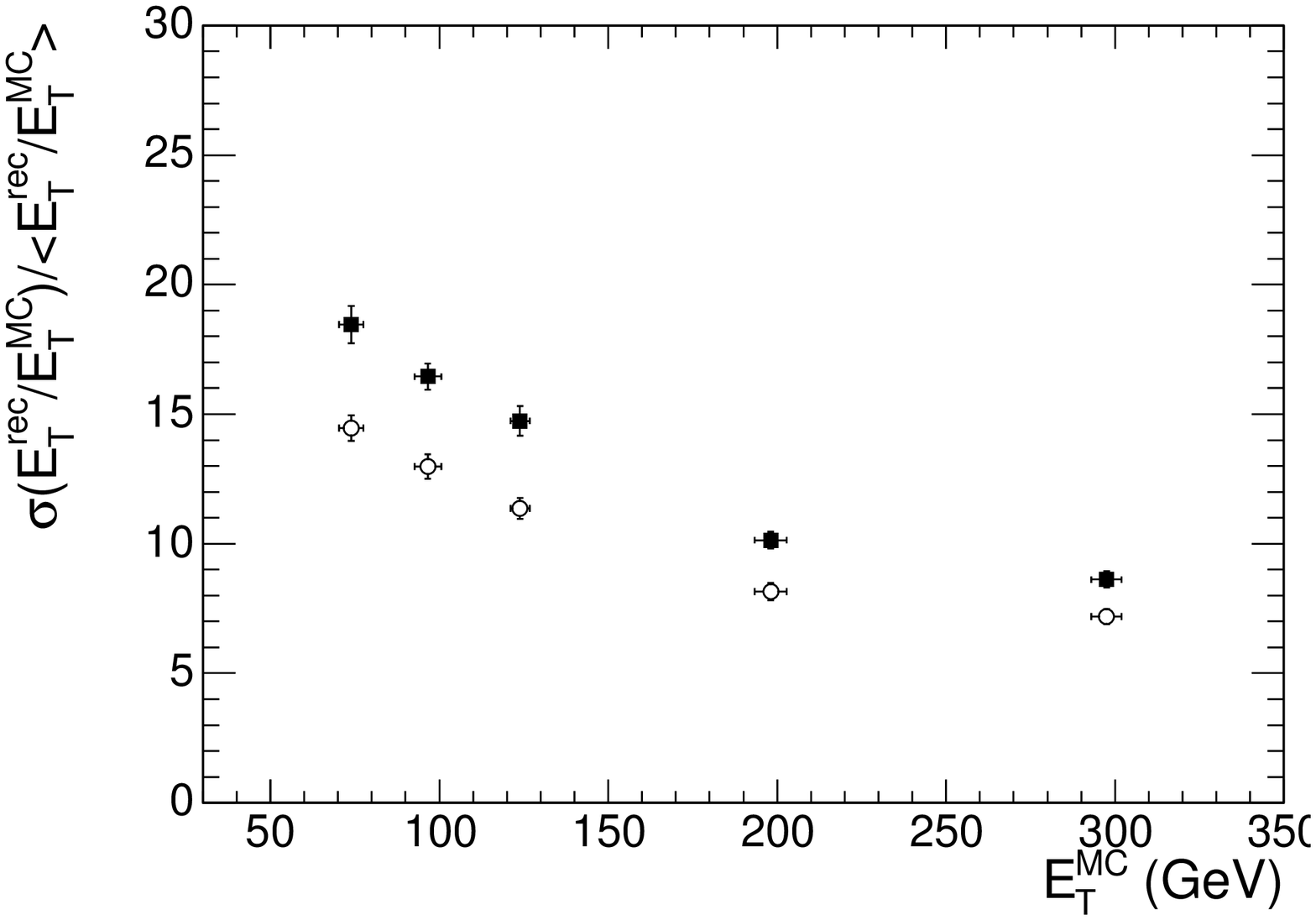}
  \caption{(Left) Jet reconstruction efficiency and purity using barrel
  calorimeters for PYTHIA-generated jets embedded in Pb+Pb events with
  $dN_{\rm ch}/dy=5000$.  (Right) The resolution of the jet E$_T$ determination in p+p
  (open symbols) and Pb+Pb (closed symbols), also with $dN_{\rm ch}/dy=5000$.}
  \label{fig:purity}
\end{figure}

The CMS calorimeters allow jet reconstruction in heavy ion 
collisions over full azimuth and a large rapidity range.
We employ  a modified iterative cone-type
jet finder that includes an event-by-event subtraction
of background energy.  
This fast method, based on calorimeter information, is available at the
trigger level and already provides excellent reconstruction efficiency and
purity, as shown in the left panel of Fig.~\ref{fig:purity}. The energy
resolution for 100~GeV jets is $\approx$ 16\% (see right panel of
Fig.~\ref{fig:purity}) and the jet location resolutions in $\eta$ and $\phi$ are
0.028 and 0.032, respectively.

\textbf{Particle identification}:
  At the LHC,  identification of particles with open and hidden heavy flavors will be 
  of primary importance. 
  The large acceptance, high-resolution muon system, in combination
  with secondary decay tagging by the silicon tracker, 
  probes the interaction of identified heavy quarks with the medium.

CMS allows dimuon reconstruction with an acceptance
spanning 4.8 units in pseudorapidity, the largest of any heavy ion
detector. 
In the central barrel, the dimuon
reconstruction efficiency is above $\sim$~80$\%$ for all
multiplicities, while the purity decreases slightly with
$dN_{\rm ch}/d\eta$ but stays above 80\% even at multiplicities as
high as $\dNdeta$ = 6500. 

\begin{figure}[htb]
\centering
\includegraphics[width=0.45\textwidth]{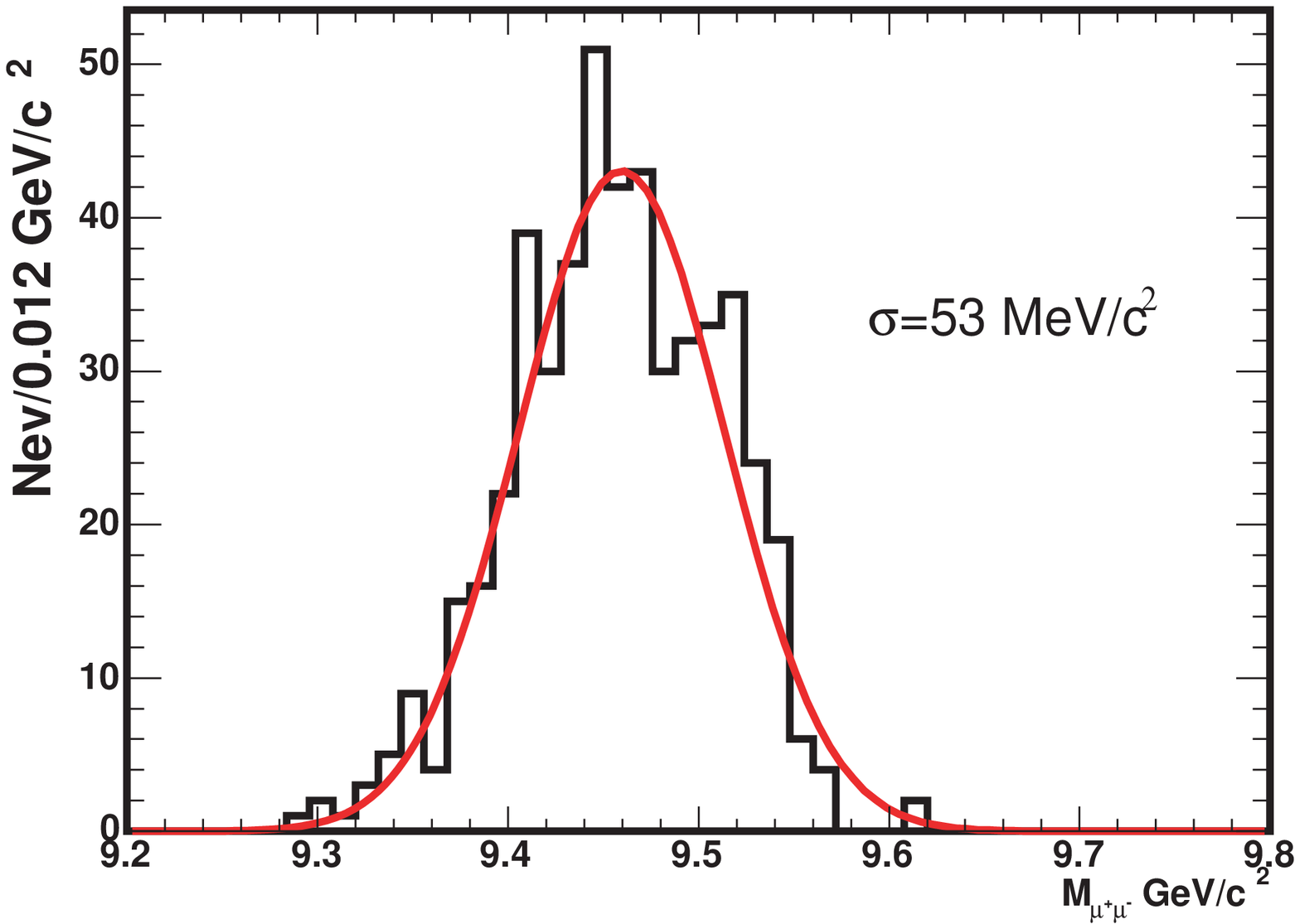}
\includegraphics[width=0.45\textwidth]{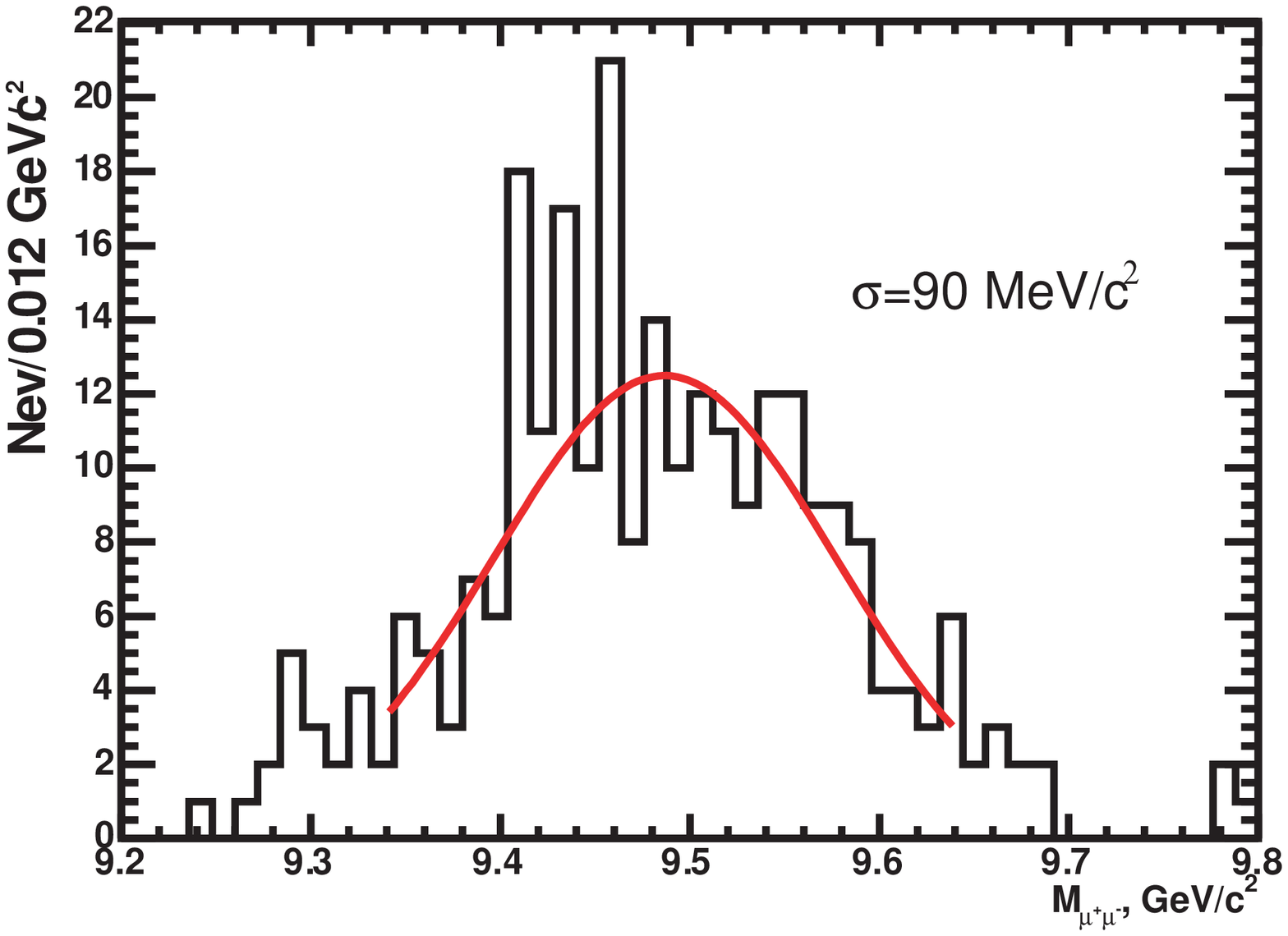}
\caption{The reconstructed $\Upsilon$ mass. (Left) Without Pb+Pb
background events and with both muons in $|\eta|< 0.8$ (barrel region
only). (Right) In a Pb+Pb
event with multiplicity $\dNdeta$=2500 and both muons in
$|\eta|<2.4$.}
\label{fig:upsilon}
\end{figure}

Figure~\ref{fig:upsilon} shows the mass resolution of the
reconstructed muon pairs from $\Upsilon$ decays obtained with  full
simulation and reconstruction. The good muon momentum resolution
translates to an $\Upsilon$ mass resolution in p+p of 53~MeV/c$^2$
in the barrel pseudorapidity region, the best of all LHC detectors,
providing a clean
separation between the members of the $\Upsilon$ family.

  The physics of
  meson vs.\ baryon production at intermediate and large $p_{T}$ can be studied using
  the results for reconstructed $\pi^{0}$s as well as the
  information provided by the silicon tracker, in combination with
  the electromagnetic and hadronic calorimeters. In the low
  transverse momentum regime, further studies will be performed to
  evaluate particle identification based on the
  specific ionization in the silicon detectors and the
  reconstruction of hadronic resonances using invariant mass
  analysis.

\section{Learning about QCD from heavy ion collisions at LHC}
\label{HIPhysics}

Heavy ion collisions at the LHC will extend our knowledge about 
strongly interacting matter at the highest energy density in two ways.
First, the energy densities of the thermalized matter are predicted
to be 20 times higher than at RHIC, implying a doubling of the initial
temperature \cite{ivanqm}.  
The higher densities of the produced partons result
in more rapid thermalization and a large increase in the time 
spent in the quark-gluon plasma phase \cite{ivanqm}.  
In addition, the abundance of hard probes will allow more detailed
quantitative measurements of the properties of the 
the initial hot and dense medium. Below we describe several of 
the physics measurements planned with CMS at the LHC.

\subsection{The Initial State and Global Observables}

Measurements at RHIC suggest that the initial state described by the concept 
of parton saturation is directly reflected in the multiplicity of produced hadrons
and their phase space distribution.  
These global features of multiparticle production, measured by PHOBOS
and others, exhibit great simplicity such as factorization into 
separate dependencies on energy and geometry and limiting fragmentation over 
a large fraction of the rapidity range. 

The large coverage of the CMS tracking detectors and the nearly full coverage of the CMS calorimeters
allows the precise determination of 
$dN_{\rm ch}/d\eta$, transverse energy (both $\ET$ and $d\ET/dy$), azimuthal anisotropy, and 
the energy of neutral spectators.
This information will be used for event categorization in various analyses. 
These
measurements will also be important for understanding the fundamental properties of particle
production in high energy collisions, following the studies performed at RHIC. 

\subsection{Thermalization and Elliptic Flow}

At RHIC, measurements of the elliptic flow of produced hadrons through their azimuthal distributions
relative to the reaction plane have become the main experimental tool addressing
the question of thermalization (or at least isotropization) in the early stage of the collision.
Comparisons to hydrodynamic calculations suggest that, at the 
highest energies, approximate thermal equilibrium
is achieved and the produced medium is characterized by a very small
shear viscosity \cite{Teaney03}. 

Measurements at the LHC will add crucial new information to the existing
studies through the measurement of flow at significantly higher initial densities.
The density dependence is particularly important since the elliptic flow data 
exhibit a steady rise in $v_2$ with particle density, 
continuing up to the highest RHIC energies. 

CMS will be able to perform these measurements with 
high precision. The highly segmented, large acceptance calorimeters allow very accurate
determination of the event plane in each event. Measurements sensitive to heavy quark flavors,
e.g.\ based on single muons not originating from the main vertex, will be performed over a 
large rapidity range and out to higher $p_T$ than accessible at RHIC. In addition, the
full suite of observables such as charged hadrons, $\pi^0$s, and jets can be studied over a 
wide kinematic range as a function of both event plane and centrality.

\subsection{Initial Temperature and Quarkonia Yields}

Measurements of the charmonium (\jpsi, \psip) and
$b\bar{b}$ (\ups, \upsp, \upspp) resonances 
provide  crucial information about the properties
of high-temperature QCD matter. 
Sequential suppression of heavy quarkonia is thought to be
one of the most direct probes of quark-gluon plasma formation.
Lattice QCD calculations indicate that
color screening dissolves the ground-state quarkonium
states, \jpsi\ and \ups, at 
$T \approx 
2T_{\rm c}$ and $4T_{\rm c}$, respectively. 
Studies of \ups\ and \upsp\ production as a function
of $p_T$ have been predicted to be sensitive to the temperature of the 
early dense medium, directly addressing one of the most important questions in our 
field \cite{vogt_psiprime}. The CMS detection capabilities for the \ups\ family in terms 
of acceptance, resolution, and statistical power make it uniquely poised to perform this 
measurement.

\subsection{Transport Properties of the Medium}

Studies of parton propagation  through the matter formed in heavy ion
collisions provide access to the transport properties of the dense medium, one 
of the key questions in heavy ion physics. Measurements of 
leading hadron production and correlations at RHIC
have already pointed to strong collective effects 
governing high $p_T$ phenomena. The new kinematic regime at LHC and 
the new probes available through precision vertexing and large coverage 
calorimetry will enable decisive studies of the medium properties.

Measurements of fully formed jets above the background
of soft hadron production require transverse energies of $E_T>50$~GeV/c,
outside the range accessible at RHIC. 
Quark jets of known energy can
be produced in reactions such as $g q \rightarrow q \gamma$~\cite{wang96} or $g q
\rightarrow q  Z^0$~\cite{kvat95}.  In these cases, the energy can be determined
since the parton energy can be tagged by
the electro-weak gauge bosons which is
unaffected by the presence of the medium. 
The high granularity, large acceptance  hadronic and electromagnetic calorimeters of the CMS
detector are well suited for these measurements. The vertexing capabilities of the 
tracking system provide additional information on the energy loss of 
heavy quarks, which will help shed light on the underlying mechanism.

\section{Summary}

In summary, measurements of hard probes in heavy ion collisions at the LHC
allow quantitative studies of the transport properties of the QCD 
medium. The much larger cross sections at the LHC will provide 
not only better statistics than at RHIC, but will also give access to qualitatively new 
observables, including fully formed jets of known energy and jets originating
from identified partons. CMS is ideally suited for these studies using 
its large acceptance, high resolution calorimetry and high precision 
tracking.  The dimuon detection capabilities
of CMS  allow precision studies of the initial medium via
sequential quarkonium suppression. 
Qualitatively new measurements  include spectra of charged hadrons 
to $p_T > 100$~GeV/c, jet $E_T$ spectra to several 100 GeV,
jet-jet correlations, tagged jets 
from heavy flavors and calibrated jets tagged via real or virtual
photons or $Z^0$'s.

\end{document}